# Towards direct neuronal current imaging via ultra-low-field MR

N. Höfner, R. Körber and M. Burghoff

Physikalisch-Technische Bundesanstalt (PTB), Abbestr. 2-12, D-10587 Berlin, Deutschland

*Abstract*: The feasibility of using ultra-low-field magnetic resonance (ULF MR) for direct neuronal current imaging (NCI) is investigated by phantom measurements. The aim of NCI is to improve the current localization accuracy for neuronal activity of established methods like electroencephalography (EEG) or magnetoencephalography (MEG) (~1 cm). A measurement setup was developed addressing the main challenge of reaching the necessary sensitivity in order to possibly resolve the faint influence of neuronal magnetic fields on $^1$H nuclear spin precession. Phantom measurements close to physiology conditions are performed simulating a specific long-lasting neuronal activity evoked in the secondary somatosensory cortex showing that the signal-to-noise ratio (SNR) of the setup needs to be further increased by at least a factor of 2.

## I. Introduction

The main motivation for the attempt to detect neuronal activity by means of ULF MR is to improve the present localization accuracy for neuronal activity of established detections methods like MEG or EEG (~1 cm) [1]. The use of very low magnetic fields in the micro-tesla range avoids the weak neuronal magnetic fields to be masked by the blood oxygenation level dependent susceptibility artifacts, which are explicitly used for the functional MRI to determine activated brain areas in high magnetic fields (tesla range) [2]. For implementing ULF MR the so called DC-mechanism is chosen for detecting the influence of long-lasting neuronal magnetic fields on the $^1$H-NMR spin dynamics. The DC-mechanism exploits that long-lasting neuronal fields superimpose the field for reading out the precessing nuclear spins, thus influencing position-dependent their Larmor frequency [3].

A long-lasting neuronal activity that could be measured via the DC-mechanism is an activity evoked on the secondary somatosensory cortex by electrostimulation of the median nerve at the wrist. MEG-data, showing a dipolar field distribution after the applied stimulation, suggest an equivalent current dipole strength of maximal 50 nAm located about 3.5 cm below the head surface [3].

## II. Methods

At the beginning of the measuring sequence a prepolarizing field (30 mT) is applied in order to generate a measurable magnetization of the $^1$H-NMR spin ensemble. By a fast change to an orthogonal oriented detection field (~10 µT) a free precession decay can be induced. During the data acquisition the long-lasting activity acts as an additional local magnetic field altering the Larmor frequency. In order to resolve the small influence of the neuronal magnetic field a reference measurement is performed and the difference signal analyzed. The field symmetry of a current dipole field leads to a cancellation effect lowering the amplitude of the difference signal [3]. This effect can be avoided by using a spatial encoding in vertical direction ensuring that the current dipole is located in the center between two slices. The phase encoding technique was chosen for encoding. The corresponding spin orientations and time traces of the different fields are displayed in figure 1.

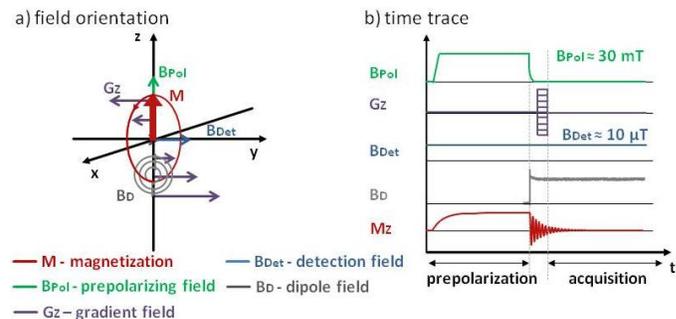

Fig. 1. a) Field orientations and b) time traces of the employed magnetic fields to detect long-lasting neuronal magnetic fields via ULF MR.

In order to measure the very weak neuronal fields within read-out fields in the micro-tesla range, magnetic shielding and an extremely sensitive sensor are necessary. For this reason ULF MR is performed in a magnetically shielded room, which consists of 2 layers of mu-metal and one eddy-current shielding and is surrounded by an additional RF shielding (see fig. 2) [3]. For signal detection a DC-SQUID current sensor connected to a first-order gradiometer is operated in a low-noise dewar. The white noise level of the sensor unit amounts to 0.5 fT/√Hz [3]. Three coils with appropriate power sources are needed for the experiment. The prepolarizing field is generated by a circular coil, which has a field-current ratio of ~2 mT/A. The biplanar gradient coil realizes a gradient-current ratio ($\partial B_y/\partial z$) of ~60 µT/Am. For ensuring a low noise level the prepolarizing and the gradient coil are disconnected from their power sources during data acquisition. As a homogeneous detection field is required during the data acquisition, a Helmholtz-type coil with a field-current ratio of ~50 µT/A in combination with a low noise current source is utilized. A head phantom with an integrated current dipole is used in order to simulate the dipolar field distribution and the time trace of the long-lasting activity. The phantom is filled with an aqueous solution containing 0.079 wt% of $CuSO_4$ to adjust the relaxation of the medium to relaxation times $T_1$ and $T_2$ of about ~100 ms [4,5] valid for grey brain tissue within micro-tesla fields. The phantom is centrally positioned below the sensor system ensuring a realistic source-sensor-distance of about 3.5 cm. The entire measurement setup shows a noise level of about 1 fT/√Hz within the used frequency range for the precession read-out. The increased noise level results from the copper of the prepolarizing coil positioned close to the sensor unit as shown in figure 2.

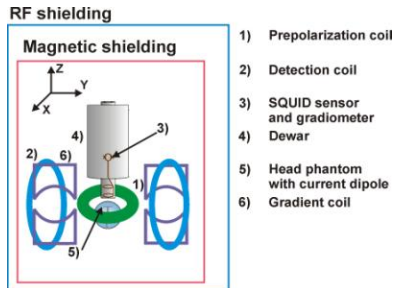

Fig. 2. ULF MR setup.

According to the adjusted relaxation times of the $CuSO_4$-solution the measurement was performed with a prepolarizing and a detection time of 0.5 s respectively. The phase encoding time amounts to 18 ms. By varying the phase gradient amplitude in increments of 2.2 µT/m, a field of view of 60 cm is obtained. The maximum gradient values of ±24.8 µT/m result in a spatial resolution of 2.5 cm. Hence, the measurement is performed with 24 phase encoding steps and 14 averages. In order not to over-heat the polarizing coil a break of 2 s is added between the measurements resulting in a total measurement time of about 45 min.

After averaging the time traces, transients which superimpose the $^1$H-precession signal and originate from the prepolarizing turn off are subtracted in the time domain. Afterwards the analytical signal is constructed before a phase correction can be applied in order to adjust the center of the FOV to the position of the phantom current dipole. A 2D-IFFT is performed and the phase difference between an influenced and a not influenced phantom measurement can be determined subsequently. Multiplying finally the phase difference picture by a magnitude image the magnitude masked phase difference is obtained.

## III. RESULTS

Figure 3 shows the magnitude masked phase difference in the different slices according to the applied current dipole strength. As the signal amplitude of the precessing spins is detected with a distance dependent sensitivity, the weak difference signals resulting from the current dipole operated within the homogeneous conductor of the head phantom occur in the slice with the smallest distance to the sensor system. The residual signal scales with the applied current dipole strength and the absence of a residual signal when the dipole was switched off demonstrates the stability of the setup. The minimal detectable current dipole strength amounts to 90 nAm (peak value).

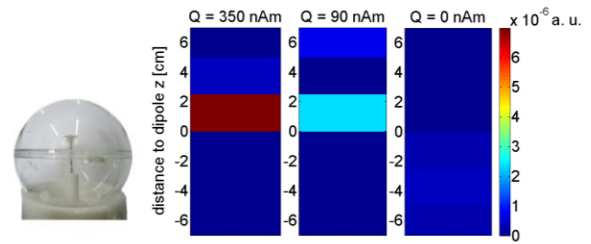

Fig. 3. Head phantom and current dipole position with respect to the magnitude masked phase differences determined for the 2.5 cm thick horizontal slices. The dewar bottom is located with a distance of 3.5 cm above the current dipole. The residual signal scales with the applied current dipole strength. If no dipole is used (0 nAm) the residual signal is below the noise level.

## IV. CONCLUSION

According to the estimated equivalent current dipole strength of about 50 nAm (peak value) for the analyzed long-lasting activity [3] the SNR of the current measurement setup needs to be improved by at least a factor of 2 to possibly realize direct neuronal current imaging via ULF MR. This SNR enhancement can be reached by further reducing the system noise level and/or increasing the prepolarizing field strength.

ACKNOWLEDGMENT

This work was supported by the Horizon 2020 Framework of the European Union (Number - 686865 - BREAKBEN).

REFERENCES

[1] C. Babiloni, V. Pizzella, C.D. Gratta, A. Ferretti, G.L. Romani, Fundamentals of Electroencefalography, Magnetoencefalography, and Functional Magnetic Resonance Imaging, International Review of Neurobiology 86, 67–80, 2009.
[2] R.H. Kraus Jr., P. Volegov, A. Matlachov, M. Espy, Towards direct neural current imaging by resonant mechanism at ultra-low field, Neuroimage 39, 310–317, 2008.
[3] R. Körber, J.O. Nieminen, N. Höfner, V. Jazbinšek. H.-J. Scheer, K. Kim, M. Burghoff, An advanced phantom study assessing the feasibility of neuronal current imaging by ultra-low-field NMR, Journal of Magnetic Resonance 237, 182-190, 2013.
[4] V.S. Zotev, A.N. Matlashov, I.M. Savukov, T. Owens, P.L. Volegov J.J. Gomez, M.A. Espy, SQUID-Based Microtesla MRI for In Vivo Relaxometry of the Human Brain. TASC 19(3), 823-6, 2009.
[5] M. Burghoff, H.H. Albrecht, S. Hartwig, R. Körber, T.S. Thömmes, H.J. Scheer, J. Voigt, L. Trahms, Squid system for meg and low field magnetic resonance. Metrol. Meas. Syst. XVI, 371-5, 2009.